\documentclass{iopart}    
\usepackage{epsfig}

\newcommand{\eqn}[1]{(\ref{eqn:#1})}

\newcommand{\doublefigure}[2]{
  \begin{center}
    \begin{tabular}{cc}
      \epsfig{file=#1, width=0.4\textwidth} &
      \epsfig{file=#2, width=0.4\textwidth} \\
      \mbox{\sl(a)} & \mbox{\sl(b)} 
    \end{tabular}
  \end{center}
  }

\begin{document} 

\title{Geodesic Deviation in Regge Calculus} 

\author{Sukanya~Chakrabarti\dag, Adrian~P~Gentle\ddag, 
  Arkady~Kheyfets\dag\ and Warner~A~Miller\ddag}

\address{\dag\ Department of Mathematics, North Carolina State
  University, Box 8205, Raleigh, NC 27695-8205, USA} 

\address{\ddag\ Theoretical Division (T-6, MS B288), Los Alamos 
  National Laboratory, Los Alamos, NM 87545, USA} 
 
\begin{abstract} 
Geodesic deviation is the most basic manifestation of the influence of
gravitational fields on matter. We investigate geodesic deviation
within the framework of Regge calculus, and compare the results with
the continuous formulation of general relativity on two different
levels. We show that the continuum and simplicial descriptions
coincide when the cumulative effect of the Regge contributions over an
infinitesimal element of area is considered. This comparison provides
a quantitative relation between the curvature of the continuous
description and the deficit angles of Regge calculus. The results
presented might also be of help in developing generic ways of
including matter terms in the Regge equations.
\end{abstract} 


\section{Introduction} 
\label{sec:I}

A generic description of the interaction between gravitational fields
and matter is an open question in the Regge formulation of general
relativity \cite{regge61,mtw,williams92}.  There has been some
progress towards the inclusion of matter terms in the Regge
formulation (cf., for instance, Dubal \cite{dubal89b}), but this has
come at the price of restrictions on the kind of matter which can be
coupled to the gravitational field, together with restrictions on the
symmetries of the field.  

A possible approach to a more general description of matter in Regge
calculus is to capitalize on our experience with the standard
continuous formulation of general relativity. If there was a
dependable way of identifying the parameters describing matter in both
descriptions one could simply insert appropriate terms in the Regge
equations, based on expressions for the energy--momentum which enters
the right hand side of Einstein's equations.

Unfortunately, such an identification is non-trivial, and
straightforward attempts on the level of a single cell (defined as the
collection of simplices attached to a single hinge in the lattice)
typically leads to divergence in the continuum limit. This is caused
by the need to compare the nonlocalizable parameters of Regge calculus
to continuum quantities localized at points, such as the
energy--momentum tensor. For some particular cases, the continuum
theory can be described using parameters which are definable within
the Regge description \cite{dubal89b}. These results, however
exciting, do not seem able to provide methods that are applicable in
more generic settings.

A better understanding of the relation between the parameters of Regge
calculus and those of the continuum description might provide some
ground for progress in this direction. In this paper, we analyze the
most basic and elementary manifestation of the influence of gravity on
matter, namely, geodesic deviation.

In Section \ref{sec:II} we briefly describe geodesic deviation in the
continuum formulation, mainly to establish the notations and
restrictions to be used in subsequent sections of the paper. In
Section \ref{sec:III} we introduce and discuss geodesic deviation in
Regge calculus. Section \ref{sec:IV} compares the Regge formulation of
geodesic deviation with the continuum description, which results in a
relation between the Riemann curvature of the continuum and the
deficit angles of Regge calculus.  Finally, in section \ref{sec:V}, we
show briefly how our results extend almost trivially to a
four-dimensional simplicial lattice.

\section{Geodesic Deviation in the Continuum} 
\label{sec:II} 

Geodesic deviation represents the simplest and most basic
manifestation of the influence of the gravitational field on
matter. It describes the relative acceleration of free test particles
(the contribution of these particles as sources of gravity is
neglected) caused by spacetime curvature. As such, it is described in
full in practically any modern text on general relativity (cf., for
instance Misner, Thorne and Wheeler \cite{mtw}, or Synge
\cite{synge}). In what follows we use notations similar to those of
Synge \cite{synge}, since they resemble more closely the expressions
we will obtain from the analysis of geodesic deviation in Regge
calculus.

\begin{figure}[t]
\centerline{\epsfig{file=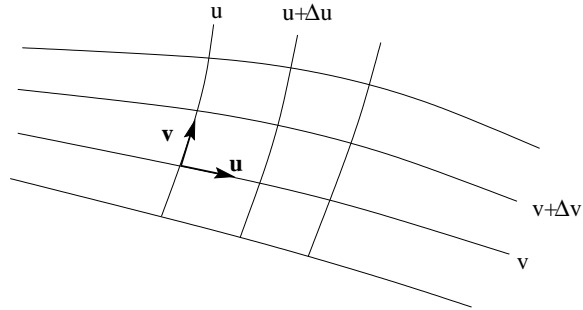,width=3.0truein}}
\caption{A one parameter family of geodesics is used to illustrate
  geodesic deviation in the continuum.  Each geodesic is identified by
  a unique label $v$, with the position along each geodesic
  parameterized by $u$. The vectors ${\mathbf u}$ and ${\mathbf v}$ are
  tangent to the $v$=constant and $u$=constant lines,
  respectively. The infinitesimal separation vector between geodesics
  is $\Delta v\, \partial_v$.}
\label{fig:continuum}
\end{figure}

The deviation of nearby geodesics is usually described by considering
a congruence of curves $C(v)$ parametrized by $v$ in such a way that
$v = {\rm const}$ along each curve, with each of these curves
parametrized by $u$. Within a coordinate patch, the whole congruence
is given by the equations $x^\mu = x^\mu(u, v)$, as indicated in
figure \ref{fig:continuum}. The curves of the congruence mesh to form
a 2--surface. That is, the vectors
\begin{equation}
{\mathbf u} = \partial_u = {\partial \over \partial u} 
\end{equation}
which is tangent to the $u$--lines (the constituent lines of the congruence),
and
\begin{equation} 
{\mathbf v} = \partial_v = {\partial \over \partial v},
\end {equation} 
which is tangent to $v$--lines, commute
\begin{equation} 
{\delta\over\delta v}\left({\partial\over\partial u}\right) =   
{\delta\over\delta u}\left({\partial\over\partial v}\right).
\end{equation} 
Here $\delta /\delta u$ and $\delta /\delta v$ are
the covariant derivatives along $u$-- and $v$--lines of the
congruence.

In studying a pair of adjacent curves $C(v)$ and $C(v + \Delta v)$ one
deals with the infinitesimal separation vector
\begin{equation} 
{\mathbf V} = \Delta v\, {\partial\over\partial v}.
\end{equation} 
Our aim is to find  how $C(v + \Delta v)$ deviates from
$C(v)$. One writes down the equation for the second covariant
derivative of the the separation vector ${\mathbf V}$, along the curve
$C(v)$ parametrized by $u$.

We are interested in the situation when the curves $C(v)$ are geodesics 
parametrized by an affine parameter $u$, 
\begin{equation} 
{\delta {\mathbf u}\over\delta u} = 0,
\end{equation} 
in which case the equation for the separation vector, called the geodesic 
deviation equation, takes the form 
\begin{equation}   
{\delta^2 {\mathbf V}\over \delta u^2} + {\mathbf R}({\mathbf u}, {\mathbf V}, {\mathbf u}) = 0.
\end{equation} 
In a coordinate frame on the patch of spacetime the equation reads 
\begin{equation} 
{\delta^2 V^\mu\over\delta u^2} + {R^\mu}_{\lambda\nu\kappa}\, 
u^\lambda V^\nu u^\kappa = 0.
\end{equation} 

In subsequent sections we will mainly have in mind the case when the
lines $C(v)$ are either timelike or spacelike, with the parameter $u$
picked to be the arc length parameter $s$,  and the $v$--lines
orthogonal to the $u$--lines (the lines $C(v)$ of the congruence). The
geodesic deviation equation then takes form
\begin{equation} 
\label{eqn:GDII}  
{\delta^2 {\mathbf V}\over \delta s^2} + {\mathbf R}({\mathbf u}, {\mathbf V}, {\mathbf u}) = 0,
\end{equation}                   
and the tangent $u$--vector ${\mathbf u}$ becomes a unit vector (the 4--velocity 
vector, if the curves $C(v)$ are timelike). 

In addition, as we shall see, the key feature of our investigation
shows up clearly for spacetimes of dimension two. We shall therefore
restrict ourselves to this case, as the exposition in higher
dimensions only clouds the basic issues. In two dimensions the Riemann
tensor has only one nonzero component, which we take to be $R^1_{\ 212}$
(cf.~Section \ref{sec:IV}).
 
\section{Geodesic Deviation in Regge Calculus} 
\label{sec:III} 

\begin{figure}[t]
  \begin{center}
    \epsfig{file=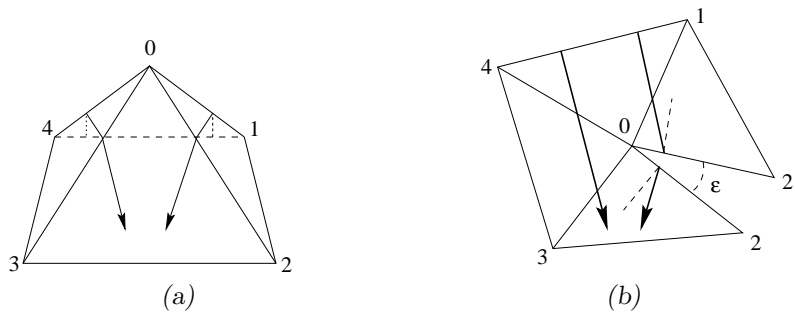,width=0.8\textwidth} \\
    \mbox{\sl(a)}  \hspace{0.4\textwidth}  \mbox{\sl(b)} 
  \end{center}
\caption{ (a) Two originally parallel test-particle world lines (dotted
  lines), are seen to converge as they enter the triangle $(023)$. The
  curvature that brings them together after the fashion of gravity on
  this two dimensional manifold is concentrated on the vertex (or
  ``hinge'') $0$, where the four triangles meet.  (b) A pictorial
  representation of the deficit angle $\epsilon$ at the vertex $0$. By
  embedding the lattice in three-dimensions, making a cut along one
  edge, and flattening the resulting structure onto a plane, the
  deficit angle at the vertex becomes apparent.  In this picture, the
  geodesics are straight lines within each triangle, and do not change
  direction as they pass from one triangle to the next.  The deviation
  of initially parallel geodesics only becomes apparent when the
  geodesics are compared within triangle $(023)$.}
\label{fig:pictorial}
\end{figure}

In this section we discuss the most elementary aspects of geodesic
deviation in Regge calculus.  We choose the pictorial representation
shown in figure \ref{fig:pictorial} to summarize the conclusions
important for our purpose.

It is sufficient to consider the case of a two--dimensional spacetime,
where the curvature is concentrated entirely at the vertices of the
simplicial lattice. At any particular vertex it is usually pictured as
the deficit angle \cite{mtw}, which emerges if one attempts to map
isometrically the neighborhood of the vertex (consisting of all
triangles emanating from the vertex) to the plane. Such a map involves
cutting the neighborhood along one or more edges coming from the
vertex and spreading it on the plane.

Figure \ref{fig:pictorial} shows two initially parallel geodesics pass
on different sides of a single vertex. The geodesics are seen to
converge (or diverge) due to the curvature concentrated at the
vertex. Each geodesic is represented by a straight line. More
precisely, the geodesic is described as being a straight line within
each triangle, and does not change direction when intersecting the
boundary of the triangle. The value and meaning of this more precise
description becomes clear when one realizes that the pictorial
representation of geodesic deviation presented in figure
\ref{fig:pictorial} is not unique.

\begin{figure}[t]
\doublefigure{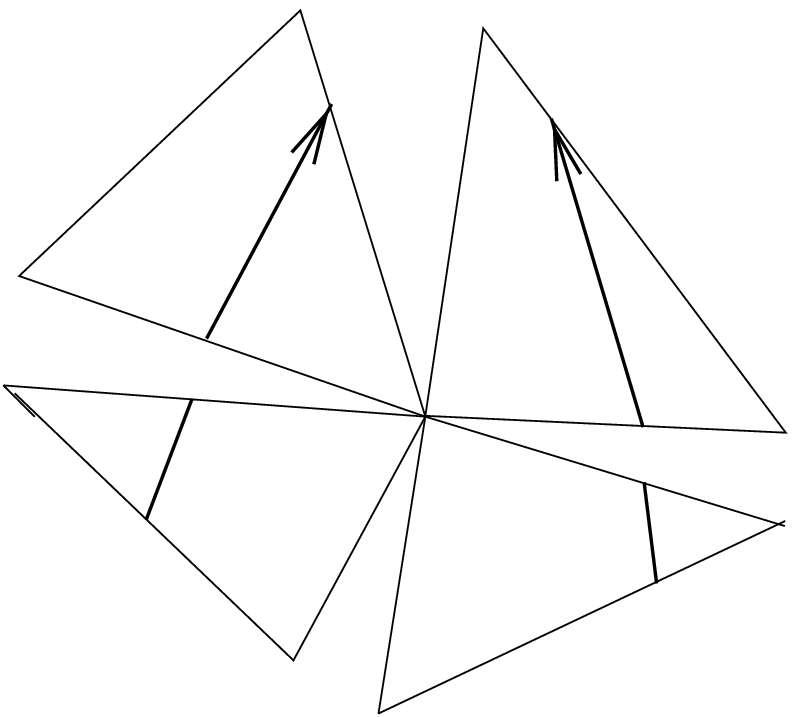}{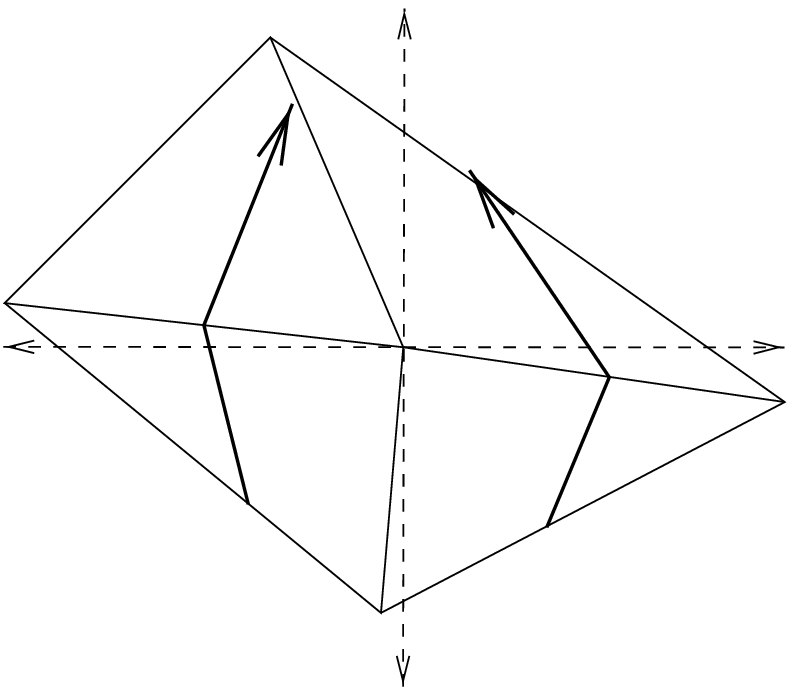}
\caption{Two other possible representations of the deficit angle about
  the central vertex. (a) A cut is made along each edge meeting at the
  vertex, and the deficit is partitioned among all such edges.
  Geodesics are now straight lines within each triangle, with pulses
  in the connection co-efficients at each triangular boundary. (b) A
  non-isometric mapping, which yields a flat metric inside each
  triangle.  The co-ordinate system is common to all triangles in the
  patch, and the metric jumps when one moves from one triangle to the
  next.  This introduces refractions in the geodesic paths.  }
\label{fig:alternative}
\end{figure}

Figure \ref{fig:alternative}(a) shows an alternative pictorial
representation, where each geodesic consists of pieces of a straight
line. The deficit angle at the vertex is represented as the sum of the
deficit angles associated with each cut. Figure's
\ref{fig:pictorial}(b) and \ref{fig:alternative}(a) both provide
truthful representations of geodesic deviation around a single vertex.

Another picture of geodesic deviation, which appears more reminiscent
of the continuum picture, can be obtained if one uses a non--isometric
map to represent the neighborhood of a vertex. For instance, one can
follow Friedberg and Lee \cite{friedberg84} and embed the neighborhood
of the vertex in a 3--dimensional space, and then project it onto a
plane (this representation is not unique, as was the case with
isometric maps).  There is no deficit angle under such a mapping. The
information about curvature is contained in the metric that changes
from one triangle projection to another (but remains constant within
each triangle). The geodesic deviation in this representation is
pictured in figure \ref{fig:alternative}(b).  Each geodesic is
represented by a piecewise straight line. It is straight within each
triangle but experiences refractions at the edges of a triangle (due
to jumps in the metric, and resulting pulses in the connection
coefficients).  For a more detailed description in a general setting,
see Williams and Ellis~\cite{williams81,williams84}.

This new picture is equivalent to that presented in figure
\ref{fig:alternative}(a), the only difference being the frame used for
describing geodesic deviation.  Figure \ref{fig:alternative}(b) uses a
holonomic frame, where all information about the curvature resides in
the metric, which changes from one triangular patch to the
next. Figure \ref{fig:alternative}(a) uses a nonholonomic frame with a
constant metric, where all information about curvature is in the
nonholonomicity object, thus generating pulses in the connection
co-efficients when crossing from one triangle to the next.

Both representations lead to a very simple description of covariant
differentiation. In the standard representation, figure
\ref{fig:alternative}(a), within each triangle we use the frame
determined by the edges emanating from a vertex, and assume that the
metric is globally flat.  Covariant derivatives inside each triangle
then coincide with partial derivatives.  When passing from one
triangle to another we perform only a transformation of the frame.

In the alternate description, figure \ref{fig:alternative}(b), the
coordinate system is common for all triangles, but the metric is flat
only within each triangle, varying from one triangle to another in a
jump.  The covariant derivative looks almost the same as in the
standard description, except that the jumps on the triangular joints
are caused by jumps in the metric.

Whatever description of geodesic deviation one chooses, for geodesics
going around one isolated vertex the curvature manifests itself in a
change of the angle of convergence of the geodesics, $\Delta\alpha$.
The total change in the angle between two such geodesics is equal to
the deficit angle $\epsilon$ associated with the vertex,
\begin{equation} 
\label{eqn:RCGD} 
\Delta\alpha = \epsilon.
\end{equation} 
Just as in the continuum case, the curvature focuses (or scatters)
geodesics.

\begin{figure}[t]
  \begin{center}
    \epsfig{file=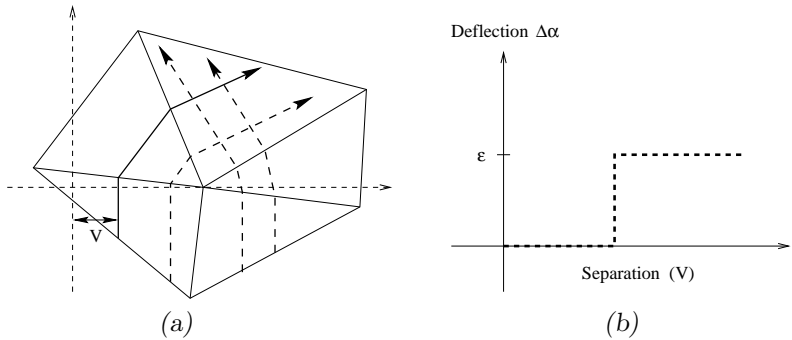,width=0.8\textwidth} \\
    \mbox{\sl(a)}  \hspace{0.4\textwidth}  \mbox{\sl(b)} 
  \end{center}
\caption{The discontinuity in the deflection angle for intially parallel test
particles as a function of the separation makes it impossible to
define, in a straighforward way, the Regge calculus analog of geodesic
deviation for an isolated vertex. (a) A collection of initially
parallel worldlines.  Two paths which pass around the same side of the
vertex do not experience geodesic deviation. (b) As a function of the
separation between the worldlines, the deviation angle shows sharp
jumps. }
\label{fig:single}
\end{figure}

This is where the similarity ends. There can be no continuous change
in $\Delta \alpha$ for a one parametric family of geodesics. That is,
geodesics on one side of the vertex remain parallel --- see figure
\ref{fig:single}.  There is no possibility of introducing, in a
straight-forward way, anything similar to the geodesic deviation
equation of the continuum, which depends on the second derivative of
the separation vector.

We are not aware of a prescription for comparing the measure of
spacetime curvature in Regge calculus with that of the continuum on
this level. Any attempt to localize curvature in this way at a single
vertex yields an inherently divergent procedure in the continuum
limit, which is not surprising.  

\section{Comparison with the Continuum: A Regge Interpretation of Geodesic 
Deviation.} 
\label{sec:IV} 

A comparison of geodesic deviation in Regge calculus with the standard
continuum description can be achieved by deriving the geodesic
deviation equation in terms of the curvature descriptors (deficit
angles) of Regge calculus, and comparing the result with the continuum
geodesic deviation equation \eqn{GDII}.
This comparison should provide a relation between the
representation of curvature in Regge calculus -- the deficit angles --
and the Riemannian curvature of the continuum.

\begin{figure}[t]
\doublefigure{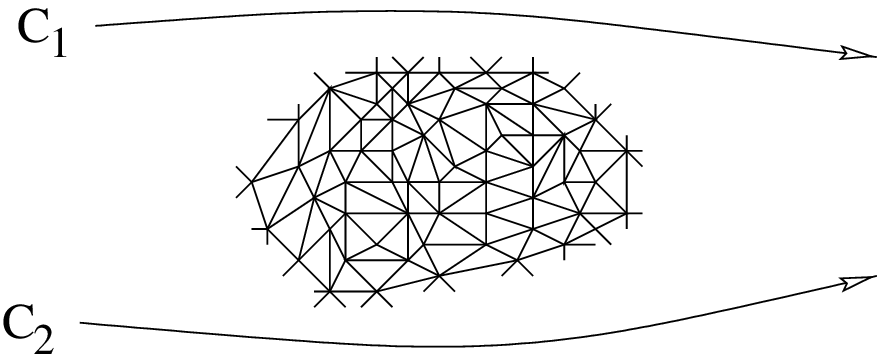}{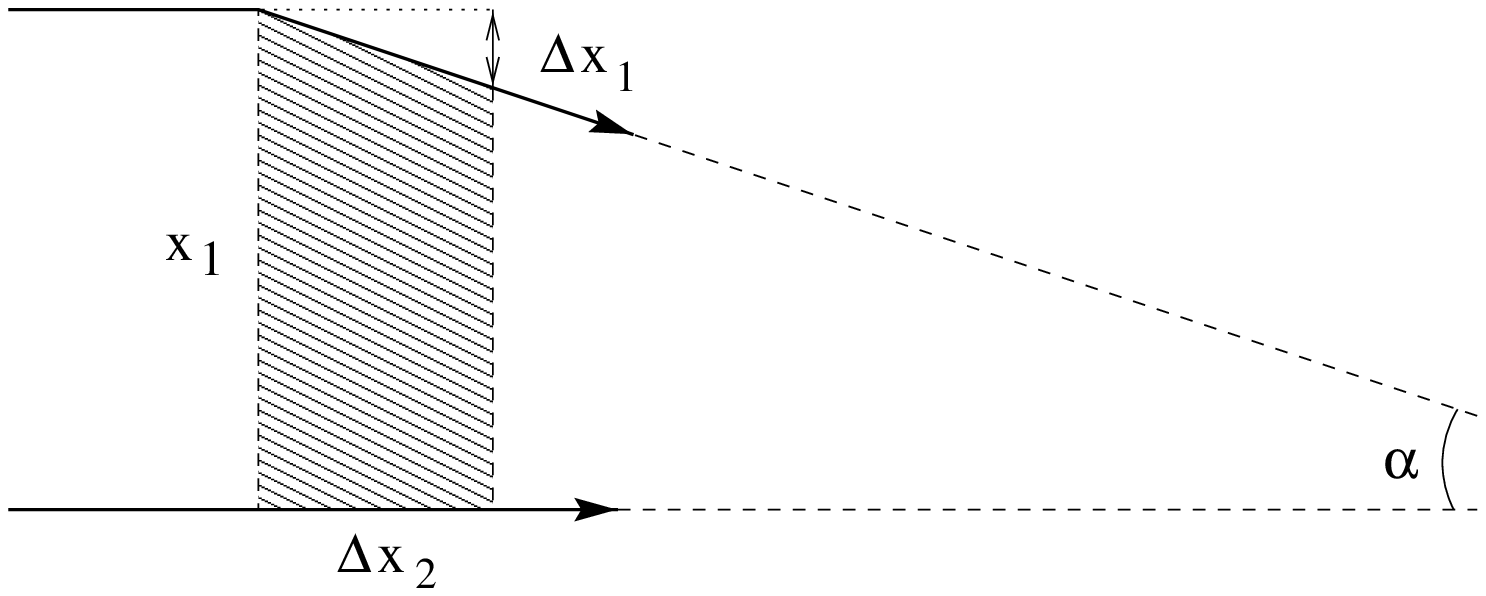}
\caption{(a) Geodesics $C_1$ and $C_2$ enclose a region containing $N$
  vertices. (b) One possible representation of the geodesic deviation
  induced in two initially parallel test particles as they pass a
  region containing $N$ vertices. }
\label{fig:deviation}
\end{figure}

In order to achieve our goal, we consider a domain of the Regge
lattice containing $N$ vertices with deficit angles ${\{
\epsilon_i\}}_{i = 1}^N$ (cf.~figure ~\ref{fig:deviation}(a)). We assume
that each deficit angle $\epsilon_i$ is small, as is the area $\Delta
A$ containing the vertices, but the total deficit angle per unit area
\begin{equation} 
\label{eqn:DPA}
\Theta = \frac{1}{\Delta A}\,{\sum_{i = 1}^N} \epsilon_i = n
\overline{\epsilon}
\end{equation} 
is finite. The symbol $n$ in equation (\ref{eqn:DPA}) represents the
density of vertices per unit area, and $\overline{\epsilon}$ is the
average deficit angle per vertex, defined as
\begin{equation} 
\overline{\epsilon} = \frac{1}{N}\,{\sum_{i = 1}^N \epsilon_i}.
\end {equation} 
The precise measure of the area is unimportant at this stage, and will
be provided later. It is important to note, however, that the total
deficit $\sum_{i = 1}^N \epsilon_i$ is small, too.

Let us consider the portion of the congruence of geodesics between the
geodesics $C_1$ and $C_2$ that contains these $N$ vertices, and no
others. It is clear that the change in the convergence angle between
these two geodesics after passing around the area is determined by the
total deficit angle attributed to the area,
\begin{equation} 
\label{eqn:GTOT}
\Delta\alpha = \sum_{i = 1}^N \epsilon_i.
\end{equation} 
We assume now that each geodesic of the congruence is parametrized by
the arc length parameter $s$, and that the curves of the congruence
are labeled by another parameter $v$ in such a way that the lines $s =
{\rm const}$ are orthogonal to geodesics. Let us approximate the whole
picture by a piece of differentiable manifold with its curvature
determined by the Riemannian tensor ${\mathbf R}$, which has only one
component in two dimensions.  Introduce on this manifold coordinates
$(x^1, x^2)$ in such a way that the $x^1$--coordinate lines 
satisfy $v = {\rm const}$, while $x^2$--coordinate lines follow the
geodesics of the congruence.

The image of the domain between the geodesics, which contains the $N$
vertices,  determines an infinitesimal rectangle $x^1$ by $\Delta
x^2$ with area $\Delta A = x^1\, \Delta x^2$. The components of the
separation vector ${\mathbf V}$ and the 4--velocity vector ${\mathbf u}$ are
given by
\begin{equation} 
{\mathbf V} = \langle x^1, 0\rangle 
\end{equation} 
and 
\begin{equation} 
{\mathbf u} = \left\langle 0, {\Delta x^2\over \Delta s}\right\rangle = 
\langle 0, 1\rangle,
\end{equation} 
where the ``size'' of the separation vector ${\mathbf V}$ is equal to
$x^1$. It can be seen from figure \ref{fig:deviation}(b) that the (small)
convergence angle $\alpha$ is expressed in terms of $\Delta x^1,
\Delta x^2$ as 
\begin{equation} 
\alpha = -{\Delta x^1\over \Delta x^2} .
\end{equation} 
With this in mind, equation (\ref{eqn:GTOT}) can be rewritten as 
\begin{equation} 
- \Delta\left({\Delta x^1\over\Delta x^2}\right) = \sum_{i = 1}^N \epsilon_i= 
n \overline{\epsilon}\, \Delta A = n \overline{\epsilon}\, x^1 \Delta x^2,
\end{equation} 
and taking into account that $\Delta x^2 = \Delta s$, this implies 
\begin{equation} 
{\Delta\over\Delta s}\left({\Delta V\over\Delta s}\right) + 
n \overline{\epsilon}\, V = 0 
\end{equation} 
where $V$ is just another notation for the ``size'' of the separation 
vector. 

The infinitesimal version of the last equation (cf.~remarks concerning
the covariant derivatives on the Regge lattice) yields the equation
for the first component of the separation vector ${\mathbf V} =
\langle V^1, 0\rangle$, namely
\begin{equation} \label{eqn:result}
{\delta^2 V^1\over \delta s^2} + n \overline{\epsilon}\, V^1 = 0.
\end{equation} 

The continuum geodesic deviation equation \eqn{GDII}, when applied to
the configuration shown in figure \ref{fig:deviation}, yields
\begin{equation} \nonumber
{\delta^2 V^1\over \delta s^2} + {R^1}_{212}\, V^1 = 0 ,
\end{equation} 
for the first component of the separation vector.  Comparison of this
continuum expression with equation \eqn{result} clearly demonstrates
that not only has the geodesic deviation equation of the continuum
been recovered, but the interpretation of the Riemannian curvature in
terms of the simplicial deficit angles has been obtained as well.
That is,
\begin{equation} 
{R^1}_{212} = n \overline{\epsilon} ,
\end{equation} 
and hence the single component of the Riemann curvature tensor may be
interpreted as the average deficit angle per unit area.

\section{The transition to four dimensions}
\label{sec:V}

We now consider the generalization of the previous results to four
dimensions.  It is important to note that the fundamental nature of
the problem does not change as we make the transition to a four
dimensional lattice.  In the Regge description, the curvature in
four-dimensions is concentrated on the triangular hinges of the
lattice. In the continuum, the four dimensional Riemann curvature
tensor has twenty independent components.  

Consider again the situation depicted in figure \ref{fig:deviation},
only now imagine that the infinitesimal region contains $N$ triangular
hinges with associated deficit angles $\{\epsilon_i\}_{i=1}^N$.  For a
test particle which lies in the plane orthogonal to one such hinge, the
tangent vector undergoes a rotation equal to the magnitude of the
deficit angle concentrated on the hinge.  This is identical to the two
dimensional case.

In general, our test particles do not lie on a plane orthogonal to the
hinges, and indeed, the hinges themselves have no overall
orientation. If we consider a vector that initially makes an angle
$\theta$ with the plane orthogonal to a single hinge, then the
component of the vector which lies in the plane will undergo a
rotation equal to the deficit angle.  This implies that the tangent
vector itself undergoes a rotation of $\epsilon \cos \theta$.

Let us again consider the congruence of geodesics between $C_1$ and
$C_2$, which encloses the region containing $N$ hinges. Let ${\mathbf u}$
be tangent to a geodesic in this congruence.  We define the average
effective deficit angle experienced by this geodesic to be
\begin{equation}
  \overline{\epsilon}_{\mathbf u} = \frac{1}{N}\sum_{i=1}^N 
  \epsilon _i \cos \theta_i,
\end{equation}
where $\theta_i$ is the initial angle between the tangent vector ${\mathbf
u}$ and the plane orthogonal to the hinge $\epsilon_i$.  

It is now clear how all components of the four-dimensional Riemann
tensor may be calculated in the continuum limit. Given two initially
parallel geodesics, the calculations of section \ref{sec:IV} are
applied, with the average deficit $\overline \epsilon$ everywhere
replaced with the average effective deficit in that direction,
$\overline \epsilon_{\mathbf u}$. The density of hinges per unit area,
$n$, is replaced with the density of hinges per unit volume.
Repeating this calculation with various choices of ${\mathbf u}$ will
yield all components of the Riemann tensor.  We see that in four
dimensions, the Riemann tensor may be viewed as the average effective
deficit angle per unit volume.

\section{Conclusion} 
\label{sec:VI} 

We have considered geodesic deviation in Regge calculus. On the level
of an elementary cell the similarity between the Regge and continuum
descriptions is rather tenuous, reduced only to the focusing (or
scattering) of geodesics. Attempts to localize the description of
geodesic deviation on this level result in a divergent procedure.

The situation considerably improves when we compute the cumulative
effect of the vertices contained in an infinitesimal element of area
(infinitesimal in the context of the continuum description). In fact,
by considering this region we reproduce the geodesic deviation
equation of the continuum, and, in doing so, obtain an interpretation
of the Riemannian curvature of spacetime as the effective deficit
angle per unit volume. The localization procedure (infinitesimal by
the continuum standard) becomes convergent in the continuum limit, and
the cause of the divergences on the level of one cell becomes, in an
obvious way, responsible for the errors of the approximation. These
may, in principle, be reduced to desirable values.
 
\ack

We are indebted to Ron Fulp, Larry Norris and Leo Brewin for many
stimulating discussions on this and related topics.  We thank Los
Alamos National Laboratory's LDRD/IP program for supporting this
research, and one of us (APG) also thanks the Centre for Non-Linear
Studies at Los Alamos National Laboratory for support.

\section*{References}

\gdef\journal#1, #2, #3, #4 {{#1}, {\bf #2}, #3 {(#4)}. }
\gdef\FP{\it{Found.~Phys.}}
\gdef\IJTP{\it{Int. J. Theor. Phys.}}
\gdef\IJMP{\it{Int. J. Mod. Phys.}}
\gdef\GRG{\it{Gen. Rel. Grav.}}

\end{document}